\documentclass[]{aastex631}

\shorttitle{Quasiperiodic Slow-Propagating EUV ``Waves"}
\shortauthors{Zhang et al. }
\graphicspath{{./}{figures/}}

\begin{document}

\title{Quasiperiodic Slow-Propagating EUV ``Wave" Trains After the Filament Eruption}

\author[0000-0001-5933-5794]{Yining Zhang}
\affiliation{State Key Laboratory of Solar Activity and Space Weather, National Astronomical Observatories, Chinese Academy of Sciences,
Beijing 100101, People's Republic of China}
\affiliation{School of Astronomy and Space Science,
University of Chinese Academy of Sciences,
Beijing 100049, People's Republic of China}

\author[0000-0001-6655-1743]{Ting Li}
\affiliation{State Key Laboratory of Solar Activity and Space Weather, National Astronomical Observatories, Chinese Academy of Sciences,
Beijing 100101, People's Republic of China}
\affiliation{School of Astronomy and Space Science,
University of Chinese Academy of Sciences,
Beijing 100049, People's Republic of China}

\author[0000-0002-5625-1955]{Weilin Teng}
\affiliation{Key Laboratory of Dark Matter and Space Astronomy, Purple Mountain Observatory, Chinese Academy of Sciences, Nanjing, Jiangsu 210023, People's Republic of China}
\affiliation{Department of Astronomy and Space Science, University of Science and Technology of China, Hefei, Anhui 230026, People's Republic of China}

\author[0000-0001-9374-4380]{Xinping Zhou}
\affiliation{College of Physics and Electronic Engineering, Sichuan Normal University, Chengdu 610068, People's Republic of China}

\author[0000-0002-9534-1638]{Yijun Hou}
\affiliation{State Key Laboratory of Solar Activity and Space Weather, National Astronomical Observatories, Chinese Academy of Sciences,
Beijing 100101, People's Republic of China}
\affiliation{School of Astronomy and Space Science,
University of Chinese Academy of Sciences,
Beijing 100049, People's Republic of China}

\author[0000-0001-5657-7587]{Zheng Sun}
\affiliation{School of Earth and Space Sciences, Peking University, Beijing 100871,
People's Republic of China}
\affiliation{Leibniz Institute for Astrophysics Potsdam, An der Sternwarte 16,
14482 Potsdam, Germany}

\author[0009-0009-7015-0024]{Xuchun Duan}
\affiliation{State Key Laboratory of Solar Activity and Space Weather, National Astronomical Observatories, Chinese Academy of Sciences,
Beijing 100101, People's Republic of China}
\affiliation{School of Astronomy and Space Science,
University of Chinese Academy of Sciences,
Beijing 100049, People's Republic of China}

\author[0000-0001-9893-1281]{Yilin Guo}
\affiliation{Beijing Planetarium, Beijing Academy of Science and Technology, Beijing 100044, China}

\author[0000-0001-8228-565X]{Guiping Zhou}
\affiliation{State Key Laboratory of Solar Activity and Space Weather, National Astronomical Observatories, Chinese Academy of Sciences,
Beijing 100101, People's Republic of China}
\affiliation{School of Astronomy and Space Science,
University of Chinese Academy of Sciences,
Beijing 100049, People's Republic of China}

\correspondingauthor{Ting Li}
\email{liting@nao.cas.cn}

\begin{abstract}
The eruption of the filament/flux rope generates the coronal perturbations, which further form EUV waves. There are two types of EUV waves, including fast-mode magnetosonic waves and slow waves. In this paper, we first report an event showing the Quasiperiodic Slow-Propagating (QSP) EUV ``wave" trains during an M6.4-class flare (SOL2023-02-25T18:40), using multiple observations from SDO/AIA, CHASE/HIS, ASO-S/FMG, SUTRI, and LASCO/C2. The QSP ``wave" trains occurred as the filament showed a rapid rise. The QSP ``wave" trains have the projected speeds of 50-130 km s$^{-1}$ on the plane of the sky, which is slower than the fast-mode magnetosonic speed in the solar corona. And the calculated period of the QSP wave trains is 117.9 s, which is in good agreement with the associated flare Quasi-Periodic Pulsation (140.3 s). The QSP wave trains could be observed during the entire impulsive phase of the flare and lasted about 30 minutes in the field of view (FOV) of SDO/AIA. About 30 minutes later, they appeared in the FOV of LASCO/C2 and propagated to the northwest. We suggest that the QSP wave trains are probably apparent waves that are caused by the successive stretching of the inclined field lines overlying the eruptive filament. The periodic pattern of the QSP wave trains may be related to the intermittent energy release during the flare.

\end{abstract}


\section{Introduction}
Large-scale wave-like global disturbance is a quite common phenomenon in the solar atmosphere. There are several types of waves including fast kink waves \citep{Aschwanden1999,Ofman2008,Guo2015}, global extreme ultraviolet (EUV) waves \citep{Zhukov2004,Patsourakos2009,Veronig2010,Li2012,Liu2012,Zhang2018,Hou2022}, and trapped sausage waves \citep{Nakariakov2003,Tian2016}. Following the successful launch of the Solar and Heliospheric Observatory (SOHO)/EUV Imaging Telescope (EIT; \citealp{Delaboudiniere1995}), EIT waves or EUV waves observed at around 195 \AA~wavelength have been a controversial topic for their exact physical nature and their relation to Moreton waves. Observations of EUV waves reveal the circularly propagating, diffuse enhancement emissions with dimming regions at EUV wavebands \citep{Moses1997,Thompson1998}. Following the launch of the Solar Dynamics Observatory (SDO; \citealp{Pesnell2012})/Atmospheric Imaging Assembly (AIA; \citealp{Lemen2012}), large-scale EUV waves also appear in 171 \AA, 193 \AA~and 211 \AA~wavebands.  

Global EUV waves are usually triggered by or accompanied with solar eruptions, especially coronal mass ejections (CMEs) \citep{Zheng2011,Zheng2012,Cheng2012,Kumar2013,Liu2018,Zhou2022b,Zhou2022a}. With the development of coronal wave theories, the exact physical nature of the coronal waves has been debated for decades. Previous studies have interpreted the waves as MHD fast-mode waves \citep{Vrsnak2000}, MHD slow-mode soliton waves \citep{Wills2007}, and non-wave models \citep{Attrill2007,Delannee2008}. In order to integrate the strengths of different models and achieve a more comprehensive interpretation of the observations, \citet{Chen2002,Chen2005} proposed a hybrid model of the EUV waves with two kinds of components. The faster component is in fact a fast MHD wave whose lower parts may refer to the coronal counterpart of the Moreton waves. While the slower component of the EUV waves is an apparent wave front caused by the successive stretching or opening of the magnetic field lines initiated by the erupting flux rope. Generally, the speed of the slow EUV waves is one-third of that of the fast ones or even less. Many previous observations have reported the fast component of the EUV waves, which has a speed of several hundreds to over 2000 km s$^{-1}$ (on the order of the speed of the fast-magnetoacoustic wave in the corona, see \citealp{Nitta2013}). The wave nature of EUV waves has been supported by the reflection, refraction, transmission \citep{Gopalswamy2009,Olmedo2012,Shen2013} and even the lensing effect \citep{Zhou2024} of EUV waves. \citet{Sun2022} provided the further observational evidence for the hybrid EUV wave model, which includes not only fast wave trains with a period of about 120 s and a velocity of $\sim$308 km s$^{-1}$, but also a wave front with a propagation velocity of about 95 km s$^{-1}$ behind the fast wave trains. The observed threefold velocity ratio between the two EUV regimes provides empirical support for the hybrid model. However, to our knowledge, the direct observations of only slow-propagating EUV waves in SDO era are rather rare.  

With the significant improvement of the observational instrument, in particular the high temporal resolution of SDO/AIA, multiple wave fronts have been reported in a single event. \citet{Liu2010} first reported the multi-ripple structures in the EUV wave accompanied by a B3.8-class flare using SDO/AIA. The multiple parallel wave fronts show a speed of 500-1200 km s$^{-1}$ and an interval of about 100 s. \citet{Liu2011} later gave the term of Quasiperiodic Fast Propagating (QFP) waves and reported the velocity over 1000 km s$^{-1}$ with an oscillation period of about 100 s. According to the numerical model of \citet{Ofman2011}, the QFP waves are intrinsically fast-mode magnetosonic waves. However, \citet{Kolotkov2021} proposed another model for QFP waves. They hold that QFP waves arise from impulsive perturbations (such as flares) evolving under waveguide-induced dispersion in coronal plasma slabs. Generally, QFP waves have speeds of over 300 km s$^{-1}$ and propagate over 500 Mm \citep{Shen2022}. Since QFP waves are tightly linked to the solar eruptions such as flares and CMEs \citep{Liu2010,Liu2011,Kumar2013,Ofman2018,Miao2019}, studying the periods is helpful in diagnosing the physical parameters of the source regions. The periods of the QFP waves range from 25-500 s in narrow QFPs (10-80\text{°}) and wide QFPs (90-360\text{°}) \citep{Liu2012,Nistico2014,Shen2018,Shen2019,Zhou2021}, and some of the QFP waves show the temporal correlations with flare Quasi-Periodic Pulsations (QPPs; \citealp{Nakariakov2007,Takasao2016,Shen2018,Zhou2022a}). 

However, according to previous studies, the multiple-ripple structures of the waves only appear in the fast-propagating scheme of EUV waves with speeds of several hundred or even over 1000 km s$^{-1}$. It remains uncertain that whether the slow-propagating EUV waves also show a multiple-ripple structure. Here, we report a unique event that contains the Quasiperiodic Slow-Propagating (QSP) EUV ``wave" trains with a period of about 118 s and speeds ranging from 50-130 km s$^{-1}$. This paper is organized as follows: Section \ref{sec:obs} presents the observational instruments used in this work and the related data analysis. Section \ref{sec:res} gives the observational results of the slow EUV wave trains. Finally, we discuss the mechanism responsible for the extremely low speed of the wave trains and the connection between the periodicity of the EUV wave trains and the flare, and give a conclusion in Section \ref{sec:dis}.

\section{Observation and Data Analysis}  \label{sec:obs}
In this study, we use data from several space-based instruments. AIA on board SDO provides high-resolution full-disk images in the EUV and ultraviolet (UV) wavebands of the corona and transition region. The spatial resolution of AIA is 0.6\arcsec per pixel and the temporal resolution is 12 s in the EUV wavebands and 24 s in the UV wavebands. The Helioseismic and Magnetic Imager (HMI; \citealp{Schou2012}) on board SDO observes the full disk photospheric line-of-sight and vector magnetograms with a pixel size of 0.5\arcsec and a temporal resolution of 45 s, respectively. The white-light observations of the CME and several subsequent waves come from the Large Angle Spectrometric Coronagraph (LASCO; \citealp{Brueckner1995}) on board the Solar and Heliospheric Observatory (SOHO), which provides the white-light images from 1.5 to 6 solar radii.

The H$\alpha$ Imaging Spectrograph (HIS) aboard the Chinese H$\alpha$ Solar Explorer (CHASE; \citealp{Li2022}) provides the raster scanning mode of the full-Sun or region-of-interest spectral images of H$\alpha$ (from 6559.7 to 6565.9 \AA)~and Fe \uppercase\expandafter{\romannumeral1} (from 6567.8 to 6570.6 \AA) with 0.024 \AA~pixel spectral resolution and 1 minute temporal resolution. The Solar Upper Transition Region Imager (SUTRI; \citealp{Bai2023}) on board the SATech-01 satellite is designed to take full disk images of the Sun at the Ne \uppercase\expandafter{\romannumeral7} 46.5 nm line with a spatial resolution of $\sim$8\arcsec and a normal temporal cadence of 30 seconds. The Full-Disk Magnetograph (FMG; \citealp{Deng2019}) on board the Advanced Space-based Solar Observatory (ASO-S; \citealp{Gan2019}) is designed to measure the full-disk vector magnetic field in the photosphere. The spatial resolution of the FMG is 1.5\arcsec, while the temporal cadence of the vector magnetogram is 2 min in normal mode and 30 s in burst mode. 

The non-linear force-free field (NLFFF) reconstruction of the coronal magnetic field is realized by using a flux rope embedding method \citep{Titov2018,Guo2019}. This method is based on the regularized Biot–Savart laws (RBSLs) proposed by \cite{Titov2018}, in which a magnetic flux rope with an axis of arbitrary shape is embedded into a potential field environment. First, we use the Line-Of-Sight magnetogram as the bottom boundary to 
perform a potential field extrapolation. Then, similar to the method used by \cite{Guo2019}, we use the triangulation of the STEREO-A and SDO 304 \AA~images to derive the filament path, which is used as the axis of the flux rope. After embedding the flux rope into the potential field model, the whole magnetic field domain is then relaxed to a force-free state by magneto-frictional relaxation \citep{Yang1986}. Finally, the reconstructed NLFFF model is re-projected into spherical coordinates to plot it over the observed image.

\section{Results}          \label{sec:res}
\begin{figure}[!htbp]
    \centering
    \includegraphics[width=0.95\textwidth]{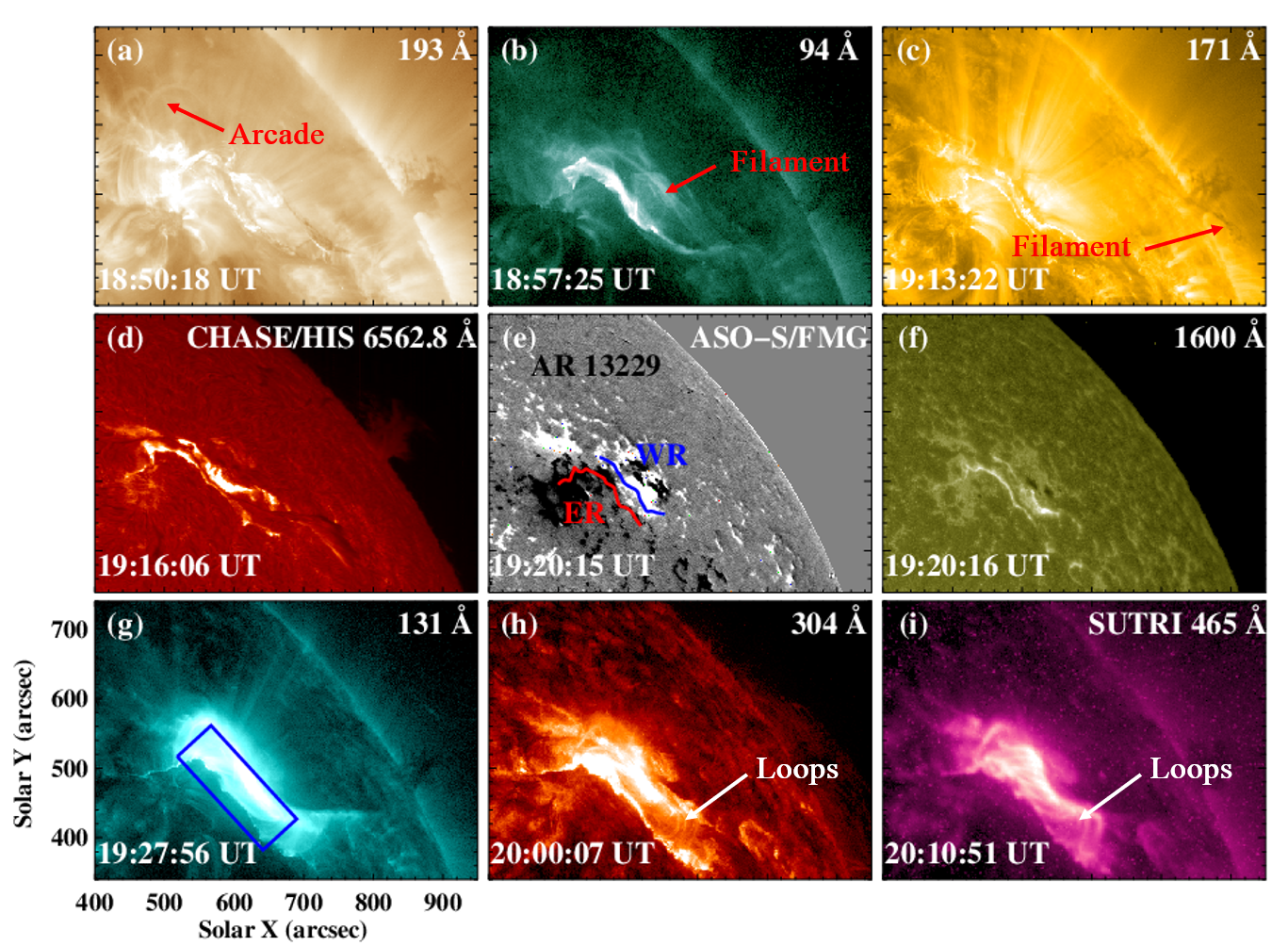}
    \caption{Overview of the M6.4-class flare SOL2023-02-25-T18:40. Panel (a): SDO/AIA 193 \AA~image with a field of view (FOV) of 550" × 400". Panels (b)-(i): SDO/AIA 94 \AA, 171 \AA, CHASE/HIS 6562.8 \AA, ASO-S line-of-sight magnetogram, AIA 1600 \AA, 131 \AA, 304 \AA, and SUTRI 465 \AA~images with the same FOV as panel (a). Red arrow in panel (a) marks the magnetic arcade above the filament. Red arrows in panel (b)\&(c) mark the filament. Red and blue lines in panel (e) mark the east ribbon (ER) and west ribbon (WR), respectively. Blue box in panel (g) denotes the position to obtain the integral intensity profiles in Figure \ref{fig:fig3}(a). White arrows in panels (h) and (i) 
   mark the post-flare loops. The animation of this figure includes FMG magnetogram, AIA 94 \AA, 131 \AA, 171 \AA, 193 \AA, and 304 \AA~images from 18:40 to 20:40 UT with a video duration of 8 s.}
    \label{fig:fig1}
\end{figure}

An M6.4-class flare occurred on February 25, 2023 in the region of NOAA active region (AR) 13229. As indicated by the GOES Soft X-ray (SXR) emission, the flare started at 18:40 UT, peaked at 19:44 UT, and finally ended at 21:00 UT. Figure \ref{fig:fig1} shows an overview of the evolution of this event with several observational instruments. Prior to the rise of the filament, there are several magnetic arcades lying over the filament (see Figure \ref{fig:fig1}(a)). The filament began to rise at 18:50 UT, stretching the surrounding arcades. The filament then underwent a rapid rise, reaching the limb of the Sun at about 19:10 UT, and finally escaping from the field of view (FOV) of SDO/AIA at about 19:20 UT. The rise of the filament could be clearly seen in Figures \ref{fig:fig1}(a)-(c). The EUV waves first appeared at about 19:05 UT at AIA 171/193 \AA~wavebands and lasted about 35 minutes until 19:40 UT. And the flare ribbons were with a configuration of a typical two-ribbon flare consisting of East Ribbon (ER) and West Ribbon (WR) (see Figures \ref{fig:fig1}(d)-(f)). The high-temperature flare loops began to form at 18:57 UT, as seen from the AIA 94/131 \AA~wavebands. And the post-flare loops appeared at almost all wavebands including AIA 171/304 \AA~and SUTRI 465 \AA~(see Figures \ref{fig:fig1}(g)-(h)). 

\begin{figure}[!htbp]
    \centering
    \includegraphics[width=0.95\textwidth]{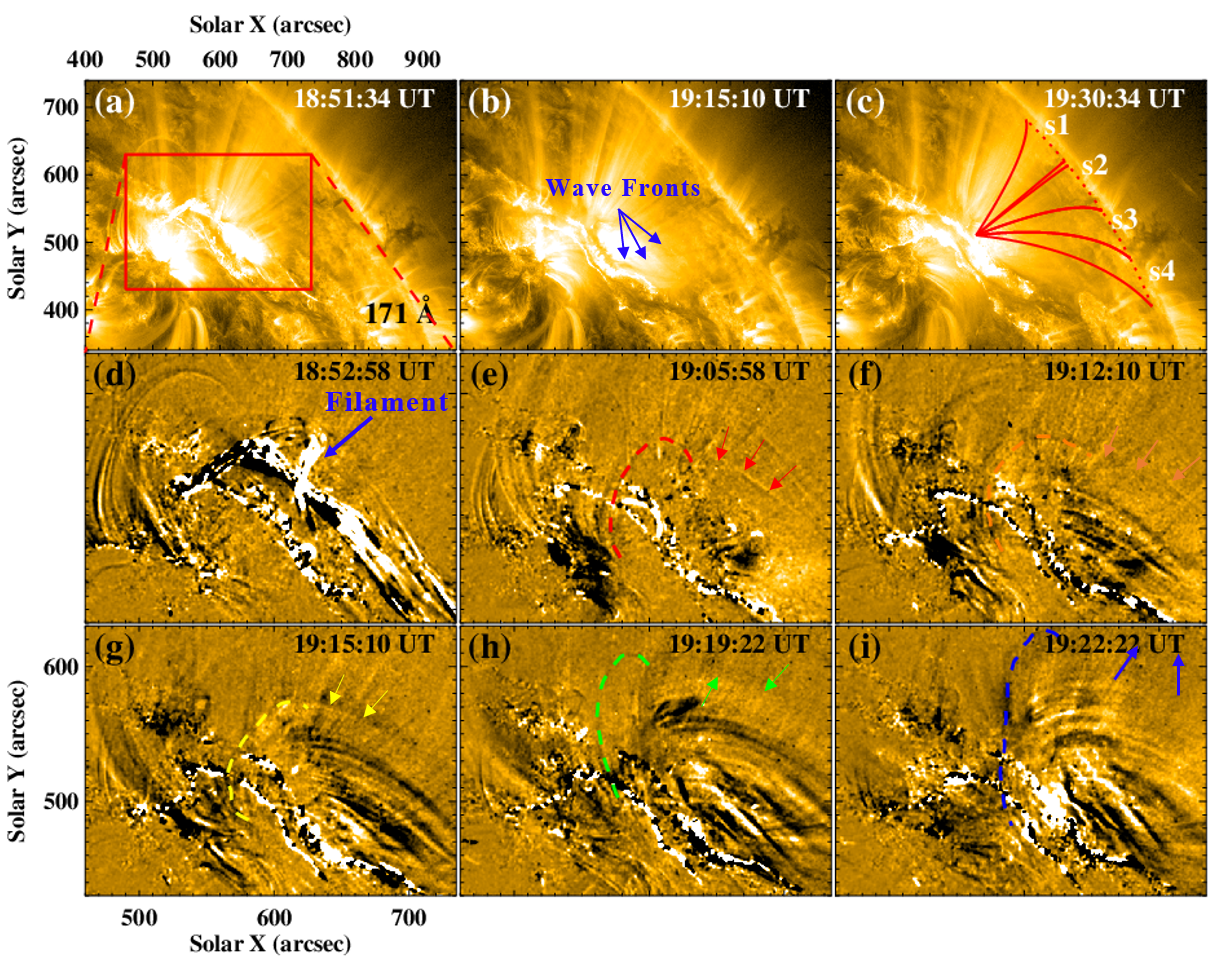}
    \caption{Propagation of EUV waves at AIA 171 \AA. Panels (a)-(c): original AIA 171 \AA~images at 18:51:34, 19:15:10 and 19:30:34 UT with the same FOV as Figure \ref{fig:fig1}, respectively. Panels (d)-(i): running difference images at AIA 171 \AA~at 18:52:58, 19:05:58, 19:12:10, 19:15:10, 19:19:22, and 19:22:22 UT with a time step of 1 minute, respectively. The red solid rectangle in panel (a) denotes the FOV of panels (d)-(i). Blue arrows in panel (b) mark the wave fronts. The red sectors `s1,' `s2,' `s3' and `s4' in panel (c) are used to obtain the stack plots at 171 \AA~in Figure \ref{fig:fig3}. Blue arrow in panel (d) indicates the filament. Colored dashed lines in panels (e)-(i) denote the inclined arcades above the filament. And colored arrows in panels (e)-(i) mark the outermost wave fronts. The animation of this figure includes 171 \AA~running difference images from 18:50 to 19:50 UT with a video duration of 15 s.}
    \label{fig:fig2}
\end{figure}

Figure \ref{fig:fig2} presents the evolution of the EUV waves at AIA 171 \AA~waveband. Before the eruption of the filament, the inclined arcades also appeared at 171 \AA~(see Figures \ref{fig:fig2}(a) and (d)). The filament began to rise gradually at about 18:50 UT and entered a fast phase at 19:06 UT. After the elevation of the filament, the waves began to appear. The dark wave fronts of the EUV waves first appeared near the WR, while the filament synchronously reached close to the solar limb (see Figure \ref{fig:fig2}(b)). From the running difference images ($\Delta$t = 1 min), we can see that the dark wave front continuously swept outward (see the colored arrows in Figures \ref{fig:fig2}(e)-(i)). The dark wave fronts appeared at about 19:05 UT near the WR (see Figure \ref{fig:fig1}(e)). As more and more EUV waves appeared, the wave trains formed and behaved an arc-shaped structure with the same direction as the expansion direction of the coronal loops until 19:35 UT. The pre-existing coronal loops were observed to expand with the locations cospatial with the outermost wave front (indicated by the colored dashed curves in Figures \ref{fig:fig2}(e)-(i)). 

\begin{figure}[!htbp]
    \centering
    \includegraphics[width=0.9\textwidth]{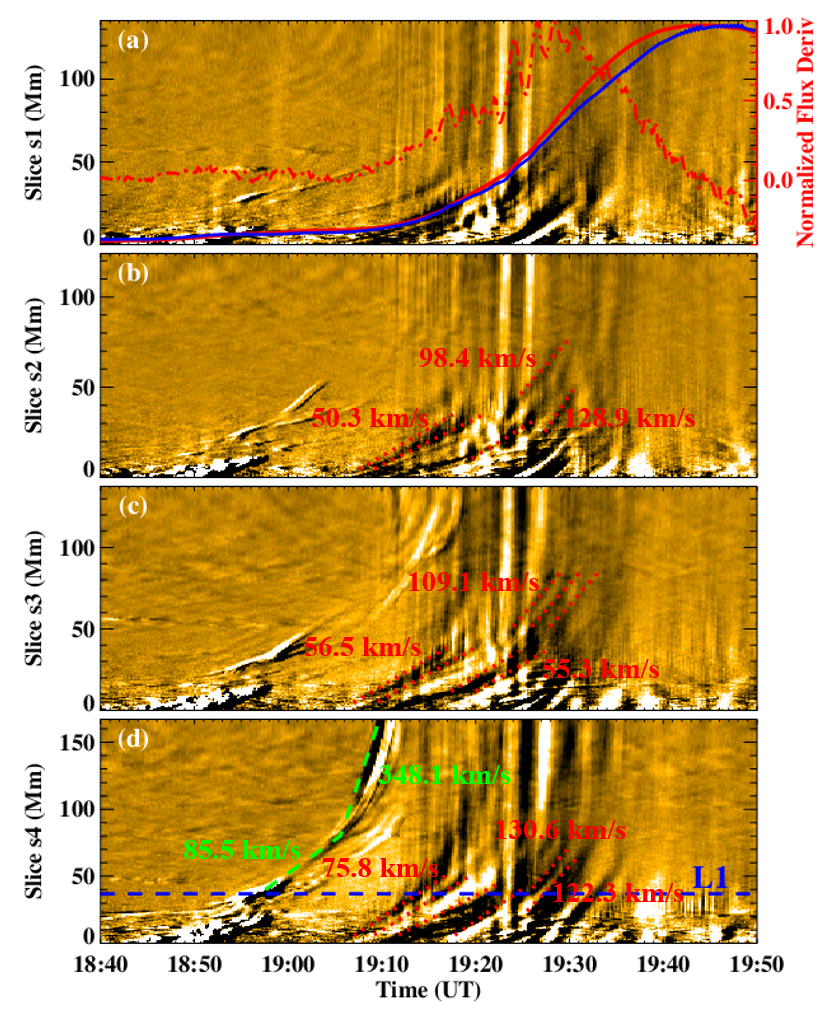}
    \caption{AIA 171 \AA~time–distance stack plots of the EUV wave trains. Panel (a): time-distance plots along sector `s1' from 18:40 to 19:50 UT. Red and blue solid lines show the normalized GOES SXR flux data at 1-8 \AA~and the normalized integral intensities in the blue box shown as Figure \ref{fig:fig1}(g) at AIA 131 \AA~waveband, respectively. The red dashed line is the normalized time derivative curve of the GOES SXR data. Panels (b)-(d): stack plots along sectors `s2', `s3', and `s4' from 18:40 to 19:50 UT, respectively. Red dotted lines in panels (b)-(d) denote the multiple wave fronts. The speeds of the waves are printed in red. The blue dashed line `L1' in panel (d) denote the position to obtain the light curve in Figure \ref{fig:fig4}(a). The greed dashed lines in panel (d) denote the gradual and fast phases of the filament elevation, respectively. The speeds of the filament are printed in green.}
    \label{fig:fig3}
\end{figure}

To investigate the characteristics of the EUV waves, we draw the stack plots in Figure \ref{fig:fig2}(c) of the sectors `s1', `s2', `s3' and `s4' from the western ribbon WR to the solar limb to obtain the time-distance plots in Figure \ref{fig:fig3} using AIA 171 \AA~running difference images. Figure \ref{fig:fig3}(a) shows the evolution of the stack plot along `s1', GOES SXR 1-8 \AA~emission (red solid line), the normalized GOES flux derivative (red dashed line), and the integral 131 \AA~intensity limited in the blue box as shown in Figure \ref{fig:fig1}(g). The two light curves show a similar trend, indicating that the SXR 1-8\AA~flux emission is mainly from the flaring kernel region. Thus, the evolution of the GOES SXR data can clearly reflect the evolution of the flare of interest. We note that there are numerous dark stripes appearing above WR especially between 19:05 and 19:35 UT. We focus mainly on the stack plots along sectors `s2' to `s4', which are closer to the wave propagation direction during 18:40-19:50 UT in Figures \ref{fig:fig3}(b)-(d). At least seven wave fronts could be distinguished from the stack plots during 19:10 to 19:30 UT. This strongly suggests that the EUV wave trains propagate from the near-ribbon across the solar surface into the interplanetary space. The calculated speeds of the wave fronts in the `s2' stack plot are 50-130 km s$^{-1}$. However, the wave trains in the `s3' and `s4' stack plots share two phases, including a slower phase with speeds of 50-70 km s$^{-1}$ accompanied by a faster phase with speeds over 100 km s$^{-1}$. From the `bright-dim' stripes, we estimate the time interval of the neighbouring wave fronts to be 2-3 minutes. The green dashed lines in Figure \ref{fig:fig3}(d) show the filament rise with two stages. First, the filament rose gradually from 18:50 to 19:06 with a speed of 85.5 km s$^{-1}$. From 19:06 UT, the filament entered a rapid rise phase with a speed of 348.1 km s$^{-1}$. We also check the stack plots with the same sectors at AIA 193 \AA~waveband. Interestingly, similar wave trains didn't appear at 193 \AA~images.

\begin{figure}[!ht]
    \centering
    \includegraphics[width=0.95\textwidth]{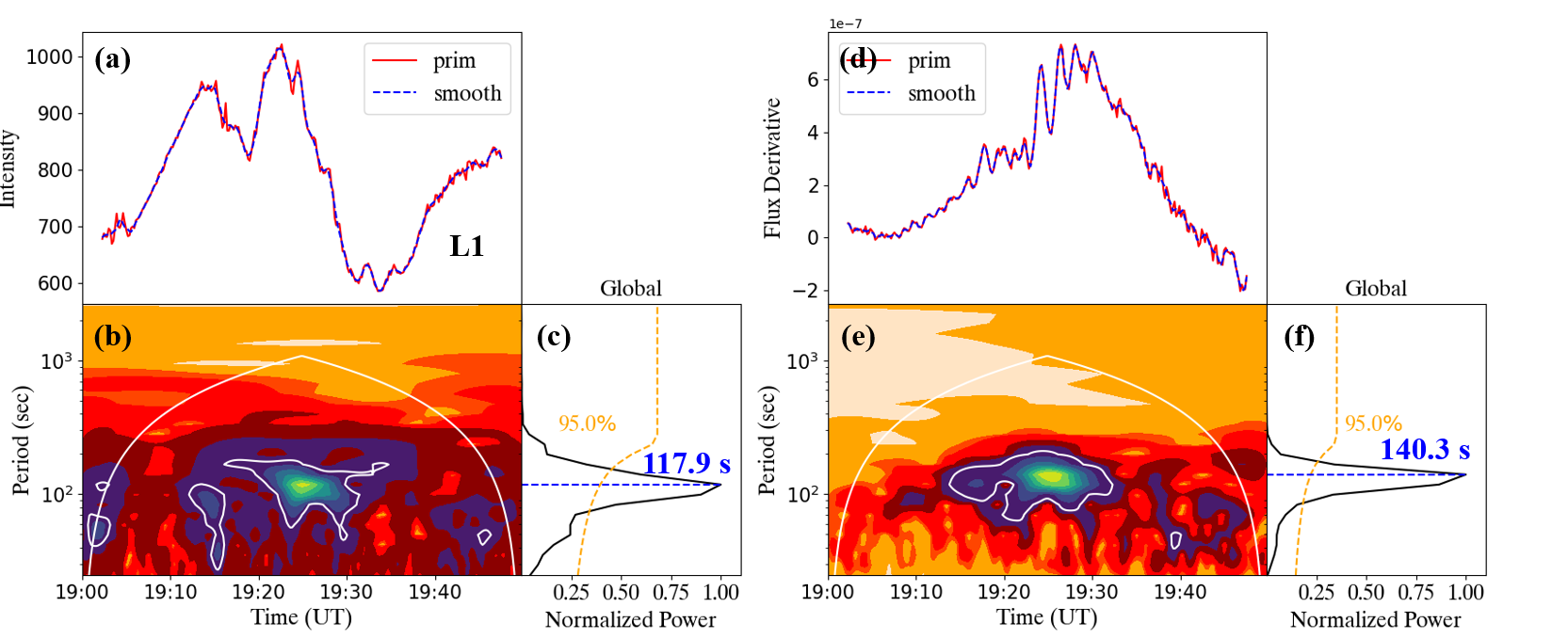}
    \caption{Wavelet analysis of the 171 \AA~light curve `L1' and GOES SXR time derivative. Panel (a): red solid and blue dashed lines indicate the original and smoothed light curves `L1' in Figure \ref{fig:fig3}(d), respectively. Panel (b): the wavelet map of the detrended intensity calculated from the difference between the two light curves in panel (a) from 19:00 to 19:50 UT. Panel (c): the corresponding global power. The orange dashed line represents the 95\% confidence level, and the blue dashed line represents the period above the 95\% confidence level with the maximum power (117.9 s). Panels (d)-(f): Similar to panels (a)-(c), but for the GOES SXR time derivative data.}
    \label{fig:fig4}
\end{figure}

From the stack plots in Figure \ref{fig:fig3}, we can see that the EUV wave trains display a quasiperiodic pattern. To further study the periodicity of the EUV wave trains, we subtract the light curve `L1' in Figure \ref{fig:fig3}(d) at 171 \AA~and perform a Morlet wavelet analysis \citep{Torrence1998} in Figure \ref{fig:fig4}. From the stack plots at 171 \AA, we estimate the time interval between neighboring stripe is 2-3 minutes with about 7 stripes in 20 minutes. To enhance the oscillations in the light curve `L1', we use the detrended light curve, which is calculated from the difference between the original light curve and smoothed light curve (red solid line and blue dashed line in Figure \ref{fig:fig4}(a), respectively), for the wavelet analysis. A smoothing window of 180 s gives a period of 117.9 s. Since the slow EUV trains occur at the impulsive phase of this flare (see the red solid line in Figure \ref{fig:fig3}(a)), we further examine the relations of the quasiperiodic EUV wave trains with the flare energy release. Wavelet analysis of the GOES SXR time derivative shows that the associated period is 140.3 s (see Figures \ref{fig:fig4}(d)-(f)). However, both curves show two pumps exceeding 95\% significance level of 117.9 s and 140.3 s, respectively. 

\begin{figure}[!htbp]
    \centering
    \includegraphics[width=0.95\textwidth]{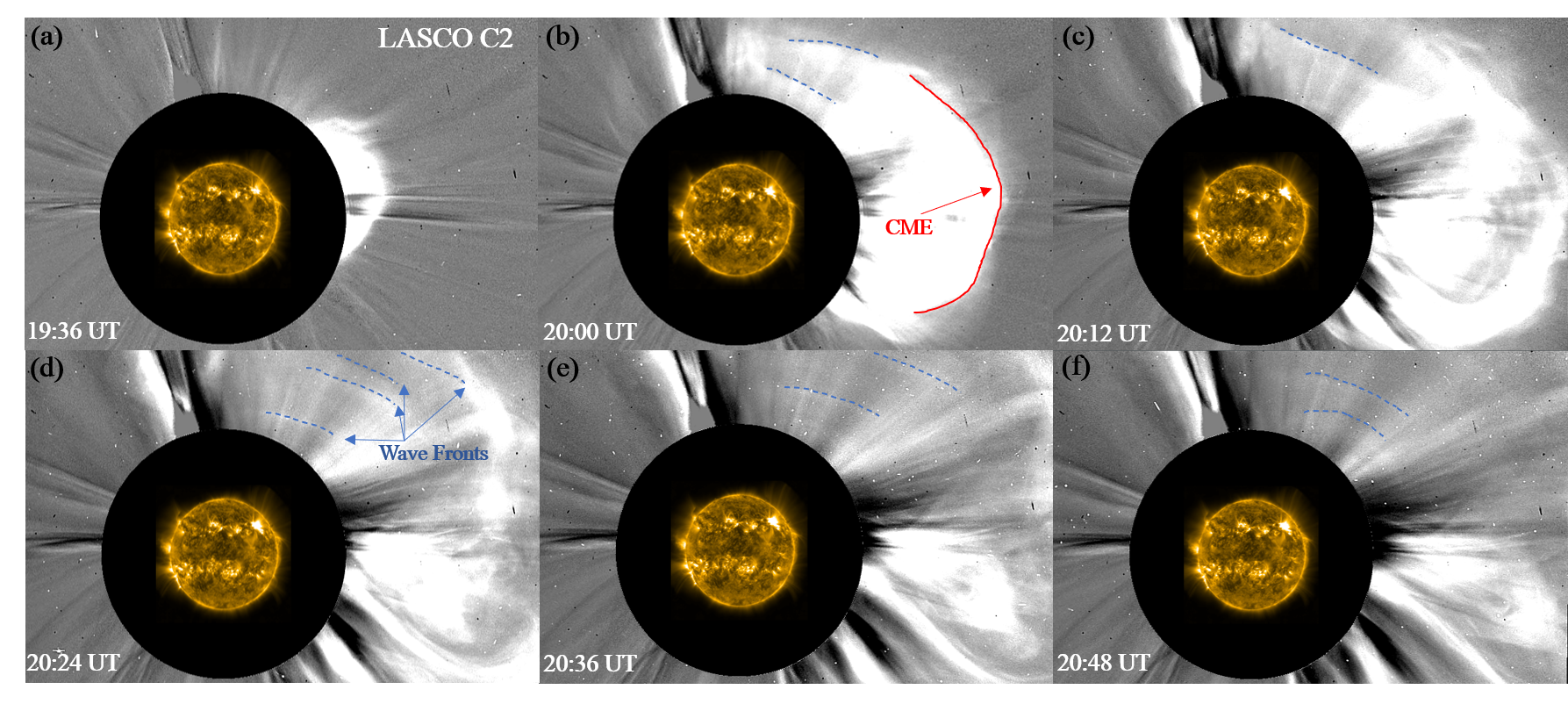}
    \caption{LASCO/C2 white-light base difference images from 19:36 UT to 20:48 UT with the time cadence of 12 minutes. Gray-blue dotted lines highlight the multiple wave fronts in the near-Sun corona. The red solid line indicates the CME. The animation of this figure includes LASCO/C2 original images from 18:00 to 23:48 UT with a video duration of 5 s.}
    \label{fig:fig5}
\end{figure}

The multiple waves were also in good agreement with observations by the LASCO/C2 white-light coronograph in the near-Sun corona. LASCO captured not only the CME caused by the eruption of the filament (see the red solid line and arrow in Figure \ref{fig:fig5}), but also multiple waves after the main CME (indicated by the gray-blue dotted lines in Figure \ref{fig:fig5}). The CME first appeared in the C2 FOV at 19:36 UT and erupted outward at 20:12 UT. The following wave fronts appear from 20:00 UT to 20:48 UT with a duration of about 48 minutes. This is in good agreement with the time sequence of the filament eruption and the EUV wave trains from 171 \AA~images.

\begin{figure}[!htbp]
    \centering
    \includegraphics[width=0.95\textwidth]{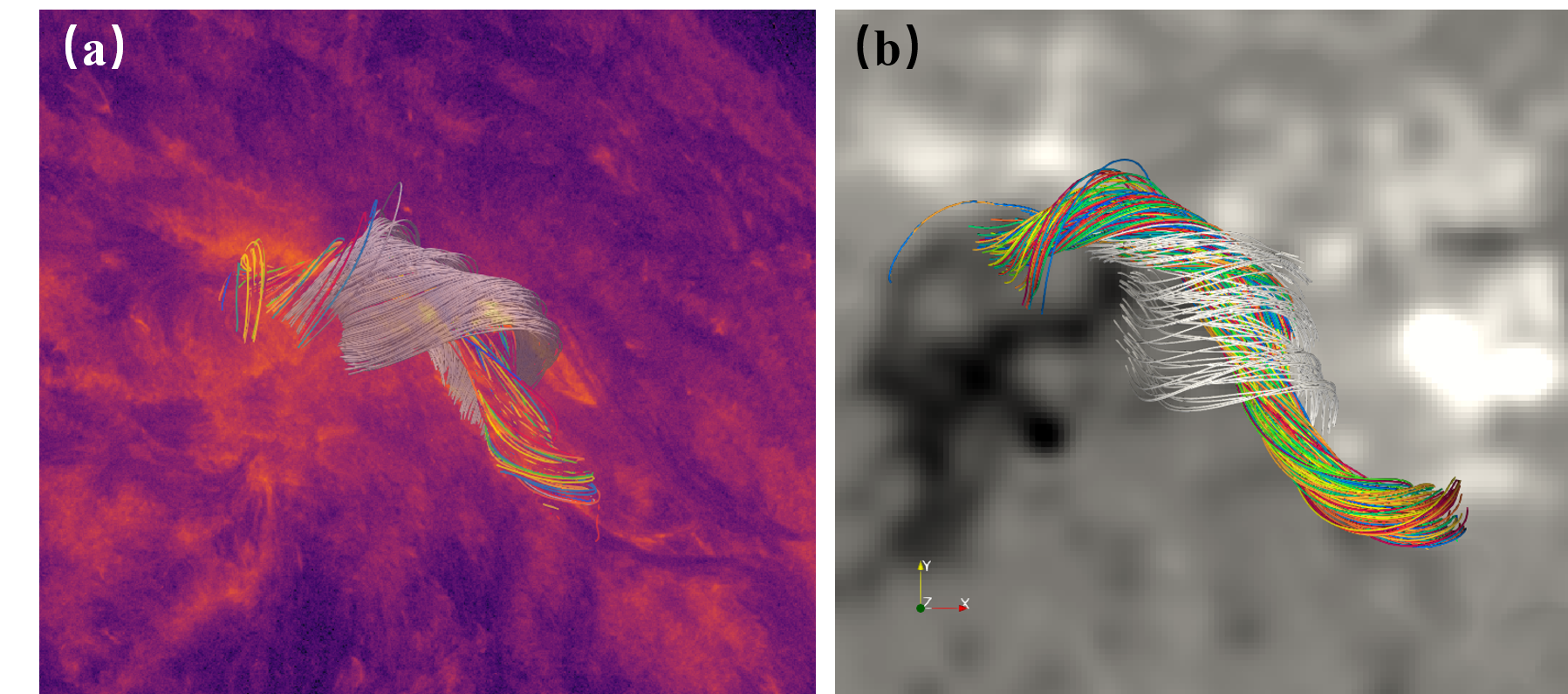}
    \caption{Reconstruction of the coronal field of the flux rope and the inclined magnetic arcades before the eruption. Panel (a): reconstruction result as seen from the Earth, superimposed on the corresponding AIA 304 \AA~image. Panel (b): top view of the reconstructed coronal magnetic field in Cartesian coordinates. The background image represents the radial magnetic field at the photosphere. White and black colors represent positive and negative magnetic polarities, respectively. Colored lines in panels (a)\&(b) represent the magnetic field lines of the flux rope, while gray lines represent the inclined arcades.}
    \label{fig:fig6}
\end{figure}

We examine the magnetic field configuration in the flare region to investigate the formation mechanism of the EUV wave trains. The reconstruction of the coronal magnetic field is performed by using a flux rope embedding method \citep{Titov2018,Guo2019}, with the HMI Line-Of-Sight magnetogram at 18:36 UT as the bottom boundary condition. The result is shown in Figure \ref{fig:fig6}. Figure \ref{fig:fig6}(a) shows the reconstructed coronal magnetic field as seen from Earth, superimposed on the corresponding AIA 304 \AA~image. The modeled flux rope is spatially consistent with the observed filament. Viewed from above, Figure \ref{fig:fig6}(b) also shows the configuration of the filament-carrying flux rope and the inclined magnetic field lines. Both results indicate that before the eruption, highly-sheared magnetic field lines tightly wrapped the filament until it erupted, causing the opening of the inclined field lines. This agrees well with the configuration of the inclined overlying arcades and the filament from Figures \ref{fig:fig1}(a) and \ref{fig:fig2}(a) at 171 and 193 \AA~wavebands. Due to the near 0-$\beta$ coronal environment, the coronal loops could be used directly to trace the magnetic field lines. Once the filament begins to rise, it will stretch the inclines field lines, which is a possible reason for the subsequent EUV waves observed after the filament eruption.

\section{Discussion and Summary}  \label{sec:dis}
The hybrid model proposed by \citet{Chen2002,Chen2005} suggests that there should be two types of EUV waves by the classification of the propagation speed. In the SOHO era, the fast component of the EUV waves has a speed of 170-350 km s$^{-1}$ and the slow component of 80 km s$^{-1}$ \citep{Dere1997,Klassen2000}. Entering the SDO era, the higher temporal resolution instrument captures the speeds of EUV waves to be over 1000 km s$^{-1}$, which is close to or even higher than the local coronal Alfv\'enic velocity. It is widely accepted that the fast EUV waves are fast MHD waves, which may be the coronal counterparts of Moreton waves \citep{Moreton1960,Smith1971,Zhang2011}. And the slow EUV ``waves", whose speed is about one-third of the corresponding fast ones, are actually the apparent propagation caused by the successive opening of the covering field lines. Here, we report a rather rare event that includes a group of slow EUV ``waves" with speeds of 50-130 km s$^{-1}$. Based on the classification by the speed, the wave trains in this study should be regarded as slow EUV ``waves" that are the apparent propagation of coronal perturbations.

\begin{figure}[!htbp]
    \centering
    \includegraphics[width=0.95\textwidth]{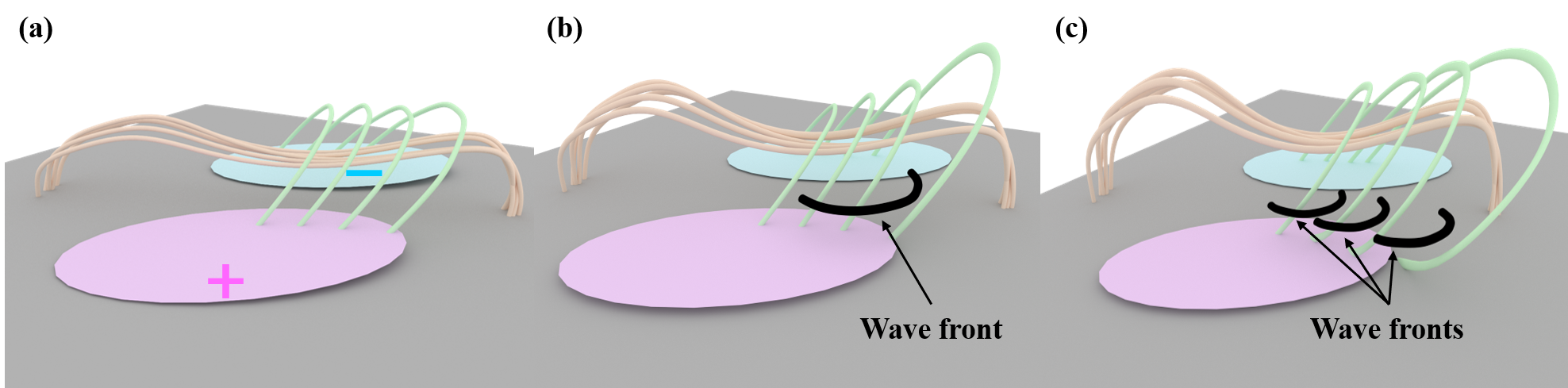}
    \caption{Schematic representations of the slow EUV wave trains. The blue and pink ellipses with minus and plus signs represent negative and positive magnetic polarities, respectively. The orange lines represent the flux rope. The green curves represent the inclined magnetic arcades above the flux rope. Thick black curves in panels (b)\&(c) represent dim wave fronts.}
    \label{fig:fig7}
\end{figure}

In Figure \ref{fig:fig7}, we give a possible scenario of the magnetic configuration for the source regions to generate slow EUV ``wave" trains. Before the eruption of the flux rope, there were multiple inclined magnetic arcades covering the filament. With the westward eruption of the filament, the inclined magnetic field lines successively opened, driving the apparent propagation of the coronal perturbations. Due to the large number of the inclined magnetic arcades, the slow EUV ``waves" propagated outward one by one, generating the QSP EUV ``wave" trains. Temporally, the occurrence of the QSP EUV ``wave" trains is consistent with the fast rise of the eruptive filament. Different from the hybrid model of \citet{Chen2002,Chen2005} and the previous observations of \citet{Liu2012} and \citet{Sun2022}, we observe multiple slow ``wave'' fronts with a period of about 2 minutes. In addition, we check the existence of QFP waves in this study. We find that there are EUV waves from 19:10 to 19:30 UT with speeds ranging from 992 to 2183 km s$^{-1}$. According to the speed classification of EUV waves, these fast components should be regarded as fast EUV waves. The wavelet analysis gives a period of about 83.4 s of the fast EUV waves. This agrees well with the hybrid model and the observations of \citet{Sun2022}. However, there are numerous studies on QFP waves since entering SDO era. So the QFP waves are not the key feature discussed in this work.

Another interesting issue in this study is the periodicity of the slow EUV ``wave" trains and their relation to the flare QPP. From Figure \ref{fig:fig4} of the wavelet analysis, the similar periodicity of the two light curves is a strong indication that the slow EUV ``wave" trains are tightly related to the flare QPPs. Some previous studies of QFP periodicity also find that some QFP waves are tightly related to flare QPPs \citep{Liu2011,Shen2012,Shen2018,Zhou2022b,Zhou2024b}. The similar periodicities suggest that the quasiperioic ``wave" trains and the quasiperiodic flare energy release have the same physical origin. When the flare releases energy intermittently, it may cause the stretching of the inclined field lines to become quasiperiodic. The newly-stretched field lines then cause the coronal perturbations with a similar period of about 2 minutes.

In this paper, we use data from SDO/AIA, CHASE, ASO-S/FMG, SUTRI and LASCO/C2 to observe the quasiperiodic slow EUV ``wave" trains accompanied by the M6.4-class flare event SOL2023-02-25T18:40. According to the AIA 171 \AA~observations, the multiple-ripple EUV ``waves" have speeds ranging from 50-130 km s$^{-1}$. This should be classified into slow-component EUV ``wave" trains, which is first reported to our knowledge. In addition, we check the periodicity of the slow EUV ``wave" trains and obtain a period of about 118 s, which is consistent with the period of GOES time derivative. This strongly suggests that the periodicity of QSP ``waves" is related to the flare QPPs. As the filament elevates, it stretches the inclined magnetic field lines and forms the apparent propagation of the EUV waves. During the impulsive of the flare, it releases energy quasiperiodically, modulating the period of the EUV ``wave" trains. 

\section*{Acknowledgments}
We thank Professor P.F. Chen at Nanjing University for the precious discussion and suggestions of this work. The authors thank the SDO/AIA, CHASE, ASO-S/FMG, SUTRI and LASCO/C2 and GOES teams for providing the data. This work is supported by the National Key R\&D Programs of China (2022YFF0503800), the National Natural Science Foundations of China (12222306, 12273060), B-type Strategic Priority Program of the Chinese Academy of Sciences (grant No. XDB0560000), and the Youth Innovation Promotion Association of CAS (2023063). We acknowledge ``Project Supported by the Specialized Research Fund for State Key Laboratory of Solar Activity and Space Weather".  

\bibliography{sample631}{}
\bibliographystyle{aasjournal}

\end{document}